\documentclass[aps, prb, twocolumn, superscriptaddress, letterpaper]{revtex4-2}

\usepackage{amsthm, amsmath, amssymb}
\usepackage{txfonts}
\usepackage{bm}
\usepackage{color}
\usepackage{float}
\usepackage{graphicx}
\usepackage[colorlinks=true, linkcolor=blue, anchorcolor=black, citecolor=blue, filecolor=black, menucolor=black, runcolor=black, urlcolor=blue]{hyperref}
\usepackage{mathrsfs} 
\usepackage{braket}
\usepackage[pagewise]{lineno}
\usepackage{orcidlink}

\newcommand{\SI}[1]{%
{\href{https://doi.org/10.1073/pnas.2422154122}{\textit{SI Appendix, }{#1}}}
}

\begin{document}
\title{Observation of disorder-induced boundary localization}

\author{Bing-Bing Wang\orcidlink{0000-0001-5119-5395}}
\thanks{These authors contribute equally.}
\affiliation{Research Center of Fluid Machinery Engineering and Technology, School of Physics and Electronic Engineering, Jiangsu University, Zhenjiang 212013, China}
\affiliation{Department of Physics, The Chinese University of Hong Kong, Shatin, Hong Kong Special Administrative Regions, China}

\author{Zheyu Cheng\orcidlink{0000-0001-5009-7929}}
\thanks{These authors contribute equally.}
\affiliation{Division of Physics and Applied Physics, School of Physical and Mathematical Sciences, Nanyang Technological University, Singapore 637371, Singapore}

\author{Hong-Yu Zou}
\affiliation{Research Center of Fluid Machinery Engineering and Technology, School of Physics and Electronic Engineering, Jiangsu University, Zhenjiang 212013, China}

\author{Yong Ge}
\affiliation{Research Center of Fluid Machinery Engineering and Technology, School of Physics and Electronic Engineering, Jiangsu University, Zhenjiang 212013, China}

\author{Ke-Qi Zhao}
\affiliation{Research Center of Fluid Machinery Engineering and Technology, School of Physics and Electronic Engineering, Jiangsu University, Zhenjiang 212013, China}

\author{Qiao-Rui Si}
\affiliation{Research Center of Fluid Machinery Engineering and Technology, School of Physics and Electronic Engineering, Jiangsu University, Zhenjiang 212013, China}

\author{Shou-Qi Yuan}
\affiliation{Research Center of Fluid Machinery Engineering and Technology, School of Physics and Electronic Engineering, Jiangsu University, Zhenjiang 212013, China}

\author{Hong-Xiang Sun\orcidlink{0000-0003-4646-6837}}
\email{jsdxshx@ujs.edu.cn}
\affiliation{Research Center of Fluid Machinery Engineering and Technology, School of Physics and Electronic Engineering, Jiangsu University, Zhenjiang 212013, China}
\affiliation{State Key Laboratory of Acoustics, Institute of Acoustics, Chinese Academy of Sciences, Beijing 100190, China}

\author{Haoran Xue\orcidlink{0000-0002-1040-1137}}
\email{haoranxue@cuhk.edu.hk}
\affiliation{Department of Physics, The Chinese University of Hong Kong, Shatin, Hong Kong Special Administrative Regions, China}
\affiliation{State Key Laboratory of Quantum Information Technologies and Materials, The Chinese University of Hong Kong, Shatin, Hong Kong Special Administrative Regions, China}

\author{Baile Zhang\orcidlink{0009-0009-7164-1933}}
\email{blzhang@ntu.edu.sg}
\affiliation{Division of Physics and Applied Physics, School of Physical and Mathematical Sciences, Nanyang Technological University, Singapore 637371, Singapore}
\affiliation{Centre for Disruptive Photonic Technologies, Nanyang Technological University, Singapore 637371, Singapore}

\begin{abstract}
Bloch wavefunctions in crystals experience localization within the bulk when disorder is introduced, a phenomenon commonly known as Anderson localization. This effect is considered universal, being applicable to all types of waves, quantum or classical. However, the interaction between disorder and topology --- a concept that has profoundly transformed many branches of physics --- necessitates revisiting the original Anderson localization picture. For instance, in the recently discovered topological Anderson insulator, the introduction of disorder induces topological boundary states that can resist localization due to protection from line-gap topology. While line-gap topology applies to both Hermitian and non-Hermitian systems, non-Hermitian systems uniquely exhibit point-gap topology, which has no Hermitian counterparts and leads to the non-Hermitian skin effect. Here, we experimentally demonstrate disorder-induced point-gap topology in a non-Hermitian acoustic crystal. This crystal, with non-Hermitian disorder in nearest-neighbor couplings, exhibits the non-Hermitian skin effect, where all eigenstates localize at a boundary. Interestingly, the boundary where localization occurs --- either the left or right --- depends on the strength of the disorder. As the disorder strength increases, the direction of boundary localization can be reversed. Additionally, we observe a ``bipolar" skin effect, where boundary localization occurs at both the left and right boundaries when disorder is introduced in next-nearest-neighbor couplings. These findings experimentally reveal a non-Hermitian mechanism of disorder-induced localization that goes beyond the conventional framework of Anderson localization.
\end{abstract}

\maketitle

\section*{Significance}
Anderson localization, first proposed by Philip Anderson in 1958, refers to the universal phenomenon of wave localization in the bulk of a crystal when disorder is introduced. Here, we experimentally demonstrate an exception using an acoustic crystal platform: adding disorder in the bulk instead causes all waves to localize at a boundary rather than in the bulk. Notably, as disorder strength increases, waves initially localized at one boundary (either left or right) shift to localize at the opposite boundary. This disorder-driven boundary localization extends beyond the conventional framework of Anderson localization and enables the use of disorder as a tuning parameter to manipulate acoustic waves, with potential applications in functional devices such as sensors.

\section*{Introduction}
Disorder plays a crucial role in understanding both quantum and classical wave phenomena in crystals \cite{anderson_absence_1958, abrahams_scaling_1979, lee1985disordered}. In the famous Anderson localization, originally proposed by Philip W. Anderson in 1958, even an arbitrarily weak random disorder is sufficient to turn all Bloch waves into localized states in the bulk, resulting in an insulator phase \cite{abrahams_scaling_1979}. Yet this property applies only to 1D and 2D systems \cite{evers2008anderson}. In 3D, the localization of all states in the bulk requires much stronger disorder \cite{lee1985disordered}, and its applicability to classical electromagnetic waves has only been numerically verified recently \cite{yamilov2023anderson}, with no successful experimental confirmation to date.

Compared to dimensionality, topology can significantly change the physical picture of Anderson localization. For instance, in the recently discovered topological Anderson insulator (TAI) \cite{li_topological_2009, groth_theory_2009, jiang2009numerical, meier_observation_2018, stutzer_photonic_2018}, the introduction of disorder opens a bandgap supporting topological boundary states, which, instead of being localized, can robustly resist localization \cite{hasan_colloquium_2010, qi_topological_2011}. This TAI concept has been extensively studied both theoretically and experimentally in Hermitian systems \cite{li_topological_2009, groth_theory_2009, jiang2009numerical, meier_observation_2018, stutzer_photonic_2018, hasan_colloquium_2010, qi_topological_2011, guo_topological_2010, mondragon-shem_topological_2014, titum_disorderinduced_2015, su2016topological, liu_topological_2020, zangeneh2020disorder, zhang_experimental_2021, cui_photonic_2022, gao_observation_2022, chen2023four} and has recently been extended to the non-Hermitian regime theoretically \cite{tang_topological_2020, zhang_nonhermitian_2020, zhang_nonhermitian_2020, liu_realspace_2021, zhang_nonhermitian_2021, luo_photonic_2022} and experimentally \cite{lin_observation_2022, gu_observation_2023, mo_imaginary-disorder-induced_2022}, for instance, by adding disordered loss and/or gain \cite{lin_observation_2022, gu_observation_2023, mo_imaginary-disorder-induced_2022}. In all these TAIs, disorder induces the so-called ``line-gap topology" \cite{hasan_colloquium_2010, qi_topological_2011, gong_topological_2018, kawabata_symmetry_2019}, which applies to both Hermitian and non-Hermitian systems. Protected by this line-gap topology, the newly generated topological states at the boundary can resist localization \cite{hasan_colloquium_2010, qi_topological_2011, stutzer_photonic_2018, tang_topological_2020}. Nevertheless, the mechanism of Anderson localization still applies to the bulk states, which become localized in the bulk by disorder.

Unlike line-gap topology, non-Hermitian systems can uniquely exhibit ``point-gap topology" with no Hermitian counterparts \cite{gong_topological_2018, kawabata_symmetry_2019}. Point-gap topology is manifested by the non-Hermitian skin effect (NHSE) \cite{gong_topological_2018, kawabata_symmetry_2019, yao_edge_2018, song_non-hermitian_2019a, borgnia_non-hermitian_2020, xiao2020non, helbig2020generalized, weidemann_topological_2020, zhang_acoustic_2021, zheng2023topological, lin_topological_2023}, where all bulk states can be localized toward a boundary. For instance, in the fundamental non-Hermitian model that achieves NHSE through asymmetric nearest-neighbor coupling, all bulk states localize in the direction along which the coupling is dominant \cite{song_non-hermitian_2019a}. The theoretical possibility of disorder-inducing point-gap topology, and thus localizing all bulk states at a boundary \cite{claes_skin_2021, zhang_nonhermitian_2021, kim_disorder-driven_2021, sarkar_interplay_2022, liu_reentrant_2023, zhang_bulk-boundary_2023}, could fundamentally change the picture of Anderson localization. However, this phenomenon has not been experimentally observed in any system  \cite{claes_skin_2021, kim_disorder-driven_2021, sarkar_interplay_2022, liu_reentrant_2023, zhang_bulk-boundary_2023}. In fact, in previously demonstrated non-Hermitian TAIs, the NHSE is suppressed rather than induced by the introduction of disorder \cite{lin_observation_2022}.

In this work, we experimentally demonstrate disorder-induced point-gap topology in a 1D non-Hermitian acoustic crystal. The non-Hermitian disorder is applied to the couplings and implemented with unidirectional acoustic amplifiers with tunable gain factors. The setup is similar to our previous demonstration of acoustic NHSE with no presence of disorder \cite{zhang_acoustic_2021}. However, in the current case, the crystal, in the clean limit (without disorder), does not exhibit NHSE. By imposing non-Hermitian disorder on the nearest-neighbor couplings, we observe NHSE, characterized by all eigenstates localized at a boundary. When the disorder strength increases, we observe the reversal of localization direction, with all eigenstates localized at the opposite boundary.  Furthermore, when the non-Hermitian disorder is imposed on the next-nearest-neighbor couplings, we observe a ``bipolar'' skin effect where localization occurs at both the left and right boundaries. Such a disorder-induced bipolar skin effect has never been previously discussed even in theory.  These results experimentally demonstrate a non-Hermitian pathway for disorder-induced localization that extends beyond the traditional understanding of Anderson localization.

\section*{RESULTS}

\subsection*{1D disordered lattice model}

\begin{figure*}
	\centering
	\includegraphics[width =  \textwidth]{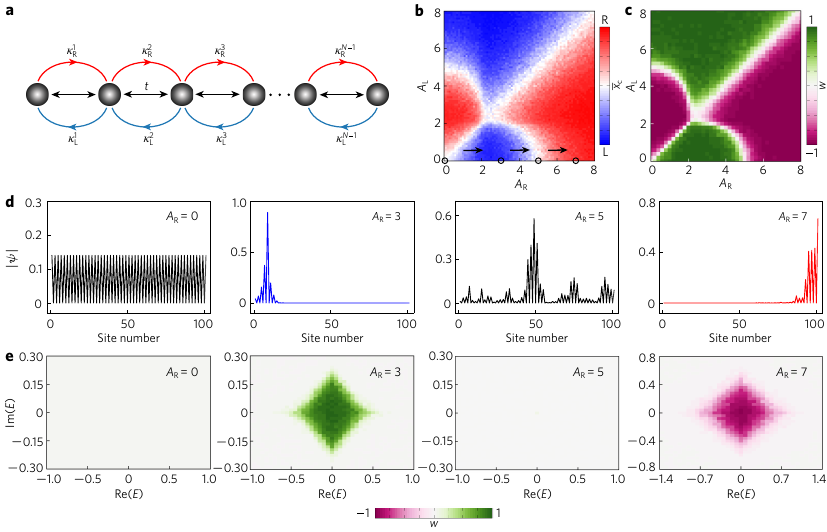}%
	\caption{ (a) Schematic of a 1D disordered lattice model with $N$ sites. (b) Calculated mean positions of eigenstates closest to zero energy. The lattice size and reciprocal coupling are $N=101$ and $t=1$, respectively. The results are averaged over 100 disorder configurations. (c) Calculated winding number against disorder strengths $A_\text{R}$ and $A_\text{L}$, with $N=2001$. The results are averaged over 50 disorder configurations. (d) Typical profiles of eigenstates closest to zero energy with increasing $A_\text{R}$ and fixed $A_\text{L}=0$, which correspond to the cases denoted by the black open circles in (b). (e) Calculated winding numbers in the complex energy plane for the four cases in (d).} 
	\label{fig01}
\end{figure*}

To facilitate the realization of disorder-induced NHSE, we consider a simplified version of the 1D tight-binding model proposed in Ref. \cite{claes_skin_2021}. As schematically shown in Fig.~\ref{fig01}(a), the lattice consists of $N$ sites, with uniform coupling (denoted by $t$) and disordered couplings (denoted by $\kappa_\text{L}^i$ and $\kappa_\text{R}^i$ for left- and right-directional couplings, respectively) between adjacent sites. Here, we set $\kappa_\text{L(R)}^i=A_\text{L(R)}r _\text{L(R)}^i$, where $A_\text{L(R)}$  is the disorder strength and $r _\text{L(R)}^i$ is a random number uniformly distributed between $-0.5$ and 0.5. The corresponding Hamiltonian reads 
\begin{equation}
 H =\sum\limits_{i=1}^{N-1} \left[(t+\kappa_\text{L}^i ) \hat c _i^{\dagger} \hat c_{i+1}+(t+\kappa_\text{R} ^i ) \hat c _{i+1}^{\dagger} \hat c_i \right],	\label{eq01}
\end{equation}
where $\hat c _i^{\dagger}$ ($\hat c _i$) is the creation (annihilation) operator at the $i$-th site. In the presence of disorder, the couplings between adjacent sites are generically nonreciprocal, i.e., $\kappa_\text{R}^i+t\neq \kappa_\text{L}^i+t$ (we assume that all couplings are real), leading to the possible point-gap topology.

To study the properties of the eigenstates, we numerically calculate the mean position of each eigenstate, defined as \cite{li_scalefree_2023}:
\begin{equation}
 x_\text{c}=\sum\limits_{j=1}^N|\psi_{j}|^2j/\sum\limits_{j=1}^N|\psi_{j}|^2,	\label{eq02}
 \end{equation}
where $|\psi_{j}|^2$ is the probability of an eigenstate at the $j$-th site. When there are skin states, the value of $x_\text{c}$ is close to 1 or $N$. We first look into states close to zero energy and will expand our interest to the states away from zero energy later. In Fig.~\ref{fig01}(b), we plot the values of $x_\text{c}$ in the $A_\text{L}$-$A_\text{R}$ plane for the states closest to zero energy, averaged over 100 independent disorder realizations. When $(A_\text{L},A_\text{R})=(0~\text{Hz},0~\text{Hz})$ (i.e., no disorder), the mean position of the state is located at the center of the lattice (white color), consistent with the characteristic of a bulk state. When disorder is activated, by contrast, we observe that for most values of disorder strengths, the states are localized around either the left (blue color) or right (red color) of the lattice. Areas in the $A_\text{L}$-$A_\text{R}$ plane with opposite localization directions are separated by sharp boundaries, indicating the phase transition.

To confirm that these localized states are induced by NHSE, we calculate the point-gap winding number from real space \cite{claes_skin_2021, borgnia_non-hermitian_2020, okuma_topological_2020, zhang_correspondence_2020}:
\begin{equation}
w(E)= \frac{1}{N}\text{Tr}(\hat{Q}^{\dagger}\lbrack\hat{Q},\hat{X}\rbrack), \label{winding}
\end{equation}
where $\hat{X}$ is the position operator. We define $\tilde Q=\mathbb{I} - 2\hat{P}$. Here, $\mathbb{I}$ is an identity matrix, and $\hat{P}$ is the projector onto the negative energy spectra of the doubled Hermitian Hamiltonian
\begin{equation}
	H_\text{D}=
	\begin{bmatrix}
		0 & H-E\\
		H^\dagger-E^*&0
	\end{bmatrix},
	\label{double}
\end{equation}
 with $E$ a reference energy in the complex energy plane. Furthermore, $\tilde Q$ can be written as~\cite{claes_skin_2021}:
 \begin{equation}
 	\tilde Q=
 	\begin{bmatrix}
 		0 & \hat Q\\
 		\hat Q^\dagger & 0
 	\end{bmatrix}.
 	\label{double}
 \end{equation} 
 Figure \ref{fig01}(c) depicts the calculated winding number in the $A_\text{L}$-$A_\text{R}$ plane at $E=0$, where the winding number takes values of $-1$, 0 and 1, corresponding to the mean positions at the right side, middle and left side of the 1D lattice model in Fig.~\ref{fig01}(b), respectively.

The phase diagram in Fig.~\ref{fig01}(c) already suggests the emergence of point-gap topology. Following the path indicated by the black arrows in Fig.~\ref{fig01}(b), for instance, we observe that a nonzero winding number is induced from a trivial system by the non-Hermitian disorder. Furthermore, multiple topological phase transitions are witnessed as the disorder strength is increased from $(A_\text{L},A_\text{R})=(0,0)$ to $(A_\text{L},A_\text{R})=(0,8)$. The evolution of the typical eigenstate closest to zero energy is plotted in Fig.~\ref{fig01}(d). When $(A_\text{L},A_\text{R})=(0,0)$, the state profile is extended as expected. By increasing $A_{\text{R}}$ to 3, the eigenstate becomes localized at the left side, consistent with $w(E=0)=1$. Then, the eigenstate distributes in the bulk of the 1D lattice again for $A_\text{R}=5$, corresponding to the phase transition point. After the phase transition, the eigenstate is localized at the right side, reflecting the change of winding number from 1 to $-1$. We also calculate the winding number in the complex energy plane for the four cases in Fig.~\ref{fig01}(d). As shown in Fig.~\ref{fig01}(e), finite regions with nontrivial winding numbers indeed appear when $(A_\text{L},A_\text{R})=(0,3)$ and $(A_\text{L},A_\text{R})=(0,7)$, proving the disorder-induced point-gap topology. The mechanism of the disorder-induced NHSE can be understood more intuitively by calculating the Lyapunov exponent (the inverse of localization length) in \SI{Sec. II}.

\subsection*{Observation of disorder-induced multiple point-gap phase transitions}

 \begin{figure*}
	\centering
	\includegraphics[width = \textwidth]{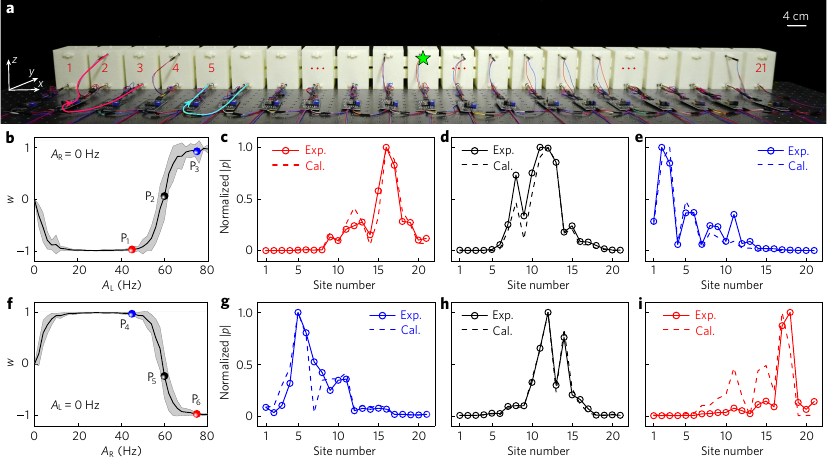}%
	\caption{(a) Photo of a 1D disordered acoustic crystal sample. The red and cyan curves indicate the externally connected sets providing positive and negative left-directional nearest-neighbor couplings, respectively. The green star denotes the sound source. (b) Calculated winding number of the 1D disordered acoustic crystal with $A_\text{R}=0~\text{Hz}$. The points $\text{P}_1$, $\text{P}_2$, and $\text{P}_3$ correspond to $A_\text{L}=45~\text{Hz}$, $60~\text{Hz}$ and $75~\text{Hz}$, respectively. (c)--(e) Experimentally measured and calculated sound pressure ($|p|$) distributions at points $\text{P}_1$ (c), $\text{P}_2$ (d) and $\text{P}_3$ (e) in (b). (f) Calculated winding number of the 1D disordered acoustic crystal with $A_\text{L}=0~\text{Hz}$. The points $\text{P}_4$, $\text{P}_5$, and $\text{P}_6$ correspond to $A_\text{R}=45~\text{Hz}$, $60~\text{Hz}$ and $75~\text{Hz}$, respectively. (g)--(i) Measured and calculated sound pressure distributions at points $\text{P}_4$ (g), $\text{P}_5$ (h) and $\text{P}_6$ (i) in (f). In the calculation in (b) and (f), we set $N=2001$, omit the intrinsic loss, and average the results over 50 disorder configurations, with the standard deviations denoted by the grey outlines. The results of (c)-(e) and (g)-(i) are obtained through a single measurement and a single calculation.}
	\label{fig02}
\end{figure*}

\begin{figure*}
	\centering
	\includegraphics[width =  \textwidth]{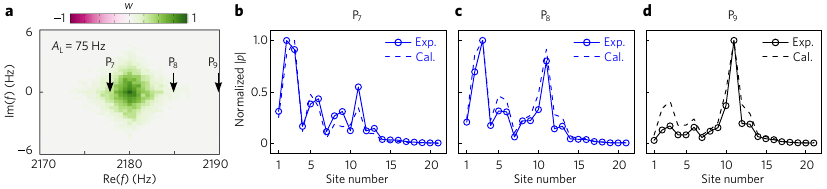}%
	\caption{(a) Calculated winding number in the complex frequency plane of the 1D disordered acoustic crystal [$(A_\text{L},A_\text{R})=(75~\text{Hz},0~\text{Hz})$]. The lattice size is $N=2001$ and the results are averaged over 50 disorder configurations. (b)--(d) Measured and calculated sound pressure ($|p|$) distributions at points $\text{P}_7$ (2178 Hz) (b), $\text{P}_8$ (2185 Hz) (c) and $\text{P}_9$ (2190 Hz) (d) in (a).}
	\label{fig03}
\end{figure*}

 To experimentally realize the lattice model in Fig.~\ref{fig01}(a), we construct a 1D acoustic crystal, whose resonators and small tubes serve as sites and reciprocal couplings of the lattice model, respectively [Fig.~\ref{fig02}(a)]. Furthermore, to realize disordered unidirectional couplings, we externally connect amplifiers to adjacent resonators through microphones (input) and speakers (output); an amplifier connected with a microphone and speaker is referred to as one external set \SI{Fig. S6(a)}~\cite{zhang_acoustic_2021}. Connection paths and gain factors of amplifiers are tuned to realize the disordered unidirectional couplings with certain strengths. Through experiments, the acoustic crystal's resonance frequency, reciprocal coupling, and intrinsic loss are obtained as $f_0=2180$ Hz, $t=-11.8$ Hz, and $\gamma_0=13$  Hz, respectively \SI{Fig. S11(a), Fig. S12(a)}. Using these parameters and the values of disordered couplings \SI{Table S4}, we can calculate the winding number and field distributions for the acoustic crystal.

Figure~\ref{fig02}(b) plots the calculated winding number as a function of $A_\text{L}$ with $A_\text{R}=0~\text{Hz}$ at 2180 Hz. Similar to the case indicated by the black path in Fig.~\ref{fig01}, three phases exist with the winding numbers $-1$,0, and 1, respectively. We select three representative points $\text{P}_1$, $\text{P}_2$, and $\text{P}_3$ in these three different phases to demonstrate the phase transition process. As shown in Figs.~\ref{fig02}(c)--\ref{fig02}(e), at $\text{P}_1$ ($\text{P}_3$), the sound field is highly localized at the right (left) boundary, showing obvious characteristics of acoustic NHSE. By contrast, the sound field has a significant distribution in the central region at point $\text{P}_2$, which is the expected behavior at the transition point. 

Furthermore, we calculate the winding number along another path of $A_\text{L}=0~\text{Hz}$ as shown in Fig.~\ref{fig02}(f). The results show opposite features compared with those in Fig.~\ref{fig02}(b). We again select three points $\text{P}_4$, $\text{P}_5$, and $\text{P}_6$, and measure the pressure amplitude distributions in the acoustic crystal. We observe that the sound field at $\text{P}_4$ and $\text{P}_6$ is localized to the opposite directions compared with $\text{P}_1$ and $\text{P}_3$, respectively. The sound field at $\text{P}_5$ is localized around the position of the sound source [Figs.~\ref{fig02}(g)--\ref{fig02}(i)]. These results reveal that, for the same amount of disorder, applying disorder to different unidirectional coupling can lead to localization in opposite directions. The small discrepancies between the experimental and calculated results is mainly due to the nonlinear effects of the amplifiers and variations in the probe insertion depth (see details in \SI{Sec. VI}).

Next, we experimentally demonstrate the frequency-dependence of the disorder-induced NHSE through the case of $(A_\text{L},A_\text{R})=(75~\text{Hz},0~\text{Hz})$. Figure~\ref{fig03}(a) depicts the corresponding winding number in the complex frequency plane. Similar to the results in Fig.~\ref{fig01}(e), the winding number is nontrivial over a range of frequencies (around 2175--2185 Hz). We select three points $\text{P}_7$, $\text{P}_8$, and $\text{P}_9$ along the real frequency axis to present the pressure amplitude profiles. At $\text{P}_7$ that is inside the nontrivial region, the sound field is localized around the left boundary [Fig.~\ref{fig03}(b)]. Near the phase transition point ($\text{P}_8$), the boundary localization becomes worse, and a pronounced peak appears at the center [Fig.~\ref{fig03}(c)]. When the excitation frequency is farther from the nontrivial region ($\text{P}_9$), acoustic NHSE disappears, and the sound field is concentrated in the central region [Fig.~\ref{fig03}(d)].

\subsection*{Observation of bipolar NHSE induced by disorder}

\begin{figure}[H]
	\centering
	\includegraphics[width =  \linewidth]{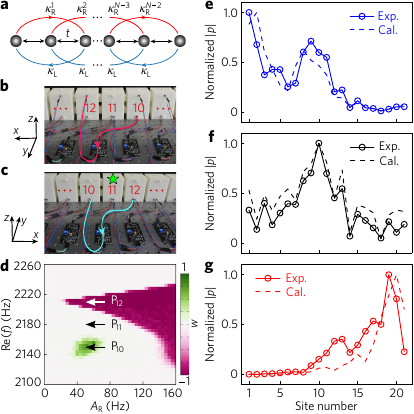}%
	\caption{(a) Tight-binding model for the bipolar NHSE, with uniform next-nearest-neighbor left-directional couplings $\kappa_\text{L}$ and disordered right-directional couplings $\kappa_\text{R}^i$. (b)-(c) Photographs of the backside (b) and frontside (c) of the fabricated sample. The green star indicates the source position. The red and cyan curves highlight the external sets serving as positive and negative next-nearest-neighbor couplings, respectively. (d) Calculated winding number of the 1D disordered acoustic crystal, with the imaginary part of the reference energy $\text{Im}(f)=0~\text{Hz}$. In the calculation, the lattice size is $N=2001$, the intrinsic loss is omitted, and the results are averaged over 50 disorder configurations. (e)--(g) Measured and calculated sound pressure distributions at points $\text{P}_{10}$ (2150 Hz) (e), $\text{P}_{11}$ (2180 Hz) (f) and $\text{P}_{12}$ (2210 Hz) (g) in (d).}
	\label{fig04}
\end{figure}

Finally, we show that disorder can induce more complicated point-gap topology such as the twisted winding topology \cite{song_non-hermitian_2019a, zhang_acoustic_2021}. Specifically, we find that certain disorders can lead to the so-called bipolar NHSE \cite{song_non-hermitian_2019a}, where skin states with opposite localizations occur simultaneously but at two frequency regions. As shown in Fig.~\ref{fig04}(a),  the lattice model is composed of $N$ sites, with the nearest-neighbor coupling $t$ and the left-directional next-nearest-neighbor coupling $\kappa_\text{L}$. The disordered right-directional next-nearest-neighbor couplings are set as $\kappa_\text{R}^i=\kappa_\text{L}+A_\text{R}r _\text{R}^i$.  The Hamiltonian of this 1D lattice model is 
\begin{equation}
H = \sum\limits_{i = 1}^{N - 1}  t\left( {\hat c_i^\dag {{\hat c}_{i + 1}} + \hat c_{i + 1}^\dag {{\hat c}_i}} \right) + \sum\limits_{i = 1}^{N - 2} \left( {{\kappa _L}\hat c_i^\dag {{\hat c}_{i + 2}} + \kappa _R^i\hat c_{i + 2}^\dag {{\hat c}_i}} \right) .
\end{equation}

To observe bipolar NHSE, we fabricate another sample [Figs.~\ref{fig04}(b)-(c)] with the following parameters: $t=-22$ Hz, $\kappa_\text{L}=-10$ Hz, $f_0=2180$ Hz, and $\gamma_0=$ 8 Hz. The corresponding phase diagram is given in Fig.~\ref{fig04}(d), where two regions with opposite winding numbers emerge within a common $A_\text{R}$ window (around $A_\text{R}=50~\text{Hz}$), exhibiting the bipolar NHSE. Note that this lattice is Hermitian and trivial when $A_\text{R}=0~\text{Hz}$, and thus this bipolar NHSE is indeed induced by disorder.

To realize desired right- and left-directional next-nearest-neighbor couplings \SI{Table S4}, we use two groups of external sets and connect them at the backside [Fig.~\ref{fig04}(b)] and frontside [Fig.~\ref{fig04}(c)] of the sample, respectively. We then put the sound source in the 11th resonator, and measure sound pressure in each resonator. Figures~\ref{fig04}(e)--(g) show the measured pressure amplitude distributions at points $\text{P}_{10}$, $\text{P}_{11}$ and $\text{P}_{12}$ in the phase diagram, corresponding to different point-gap phases with fixed $A_\text{R}=50~\text{Hz}$ [Fig.~\ref{fig04}(d)]. As can be seen, the pressure amplitude distributions at points $\text{P}_{10}$, $\text{P}_{11}$ and $\text{P}_{12}$  are localized around the left boundary, source position, and right boundary, respectively, clearly revealing the bipolar NHSE.

\section*{DISCUSSION}

In conclusion, we have demonstrated the disorder-induced NHSE in a 1D disordered acoustic crystal. The boundary localization induced by disorder presents a different scenario compared to conventional Anderson localization in the bulk. Given that NHSE exhibits a funneling effect \cite{weidemann_topological_2020,zhang_acoustic_2021} that allows unidirectional acoustic pulse transmission \cite{zhang_acoustic_2021}, our current findings further enhance the potential of NHSE in various applications such as surface acoustic wave (SAW)-based signal processing and sensing  \cite{mandal2022surface}, even within disordered systems. Disorder, inherent in practical systems, may serve as a control parameter to deliberately engineer NHSE in these applications. While our study focuses on acoustic waves, similar principles could inspire demonstrations in other wave types, including quantum systems. Extending disorder-induced boundary localization to higher-dimensional and nonlinear systems would also be an intriguing avenue for future research.

\section*{METHODS}

\bigskip
\noindent\textbf{Sample details.}---All samples are fabricated via the 3D printing technique with photosensitive resin. The thickness of the samples is 5 mm. In the sample shown in Figs.~\ref{fig02} and~\ref{fig04}, the hole with a radius of 2 mm on the top surface of each resonator is for detection. The small (radius = 3.5 mm) and big holes (radius = 2 mm) on both front and back sides are used for inserting the microphone and loudspeaker of each external set, respectively. The other small hole of the 11th resonator is used for excitation.

\bigskip
\noindent\textbf{Experimental set-up.}---In experiments, a broadband sound signal ranging from 1800--2500 Hz is launched from a balanced armature speaker (radius = 1 mm) driven by a power amplifier and guided into the samples through a narrow tube (radius = 1.5 mm). A microphone (B\&K type-4182) is used to detect sound signals.

To measure the acoustic pressure profiles along 1D acoustic crystals, the sound source is placed in the 11th resonator [green star in Figs.~\ref{fig02}(a) and ~\ref{fig04}(c)], and the microphone is inserted into the periodic holes on the top surface of each resonator to collect sound signals. The pressure profiles are retrieved by recording the measured pressure magnitude and phase through the software PULSE Labshop.

\bigskip
\noindent\textbf{Data, Materials, and Software Availability.}---The code and experimental data presented in all figures are available at \href{https://doi.org/10.21979/N9/NAQAYC}{https://doi.org/10.21979/N9/NAQAYC}. Other study data are included in the article and/or {\href{https://doi.org/10.1073/pnas.2422154122}{\textit{SI Appendix}}}.


\begin{thebibliography}{50}%
\makeatletter
\providecommand \@ifxundefined [1]{%
 \@ifx{#1\undefined}
}%
\providecommand \@ifnum [1]{%
 \ifnum #1\expandafter \@firstoftwo
 \else \expandafter \@secondoftwo
 \fi
}%
\providecommand \@ifx [1]{%
 \ifx #1\expandafter \@firstoftwo
 \else \expandafter \@secondoftwo
 \fi
}%
\providecommand \natexlab [1]{#1}%
\providecommand \enquote  [1]{``#1''}%
\providecommand \bibnamefont  [1]{#1}%
\providecommand \bibfnamefont [1]{#1}%
\providecommand \citenamefont [1]{#1}%
\providecommand \href@noop [0]{\@secondoftwo}%
\providecommand \href [0]{\begingroup \@sanitize@url \@href}%
\providecommand \@href[1]{\@@startlink{#1}\@@href}%
\providecommand \@@href[1]{\endgroup#1\@@endlink}%
\providecommand \@sanitize@url [0]{\catcode `\\12\catcode `\$12\catcode
  `\&12\catcode `\#12\catcode `\^12\catcode `\_12\catcode `\%12\relax}%
\providecommand \@@startlink[1]{}%
\providecommand \@@endlink[0]{}%
\providecommand \url  [0]{\begingroup\@sanitize@url \@url }%
\providecommand \@url [1]{\endgroup\@href {#1}{\urlprefix }}%
\providecommand \urlprefix  [0]{URL }%
\providecommand \Eprint [0]{\href }%
\providecommand \doibase [0]{https://doi.org/}%
\providecommand \selectlanguage [0]{\@gobble}%
\providecommand \bibinfo  [0]{\@secondoftwo}%
\providecommand \bibfield  [0]{\@secondoftwo}%
\providecommand \translation [1]{[#1]}%
\providecommand \BibitemOpen [0]{}%
\providecommand \bibitemStop [0]{}%
\providecommand \bibitemNoStop [0]{.\EOS\space}%
\providecommand \EOS [0]{\spacefactor3000\relax}%
\providecommand \BibitemShut  [1]{\csname bibitem#1\endcsname}%
\let\auto@bib@innerbib\@empty
\bibitem [{\citenamefont {Anderson}(1958)}]{anderson_absence_1958}%
  \BibitemOpen
  \bibfield  {author} {\bibinfo {author} {\bibfnamefont {P.~W.}\ \bibnamefont
  {Anderson}},\ }\bibfield  {title} {\bibinfo {title} {Absence of diffusion in
  certain random lattices},\ }\href {https://doi.org/10.1103/PhysRev.109.1492}
  {\bibfield  {journal} {\bibinfo  {journal} {Phys. Rev.}\ }\textbf {\bibinfo
  {volume} {109}},\ \bibinfo {pages} {1492} (\bibinfo {year}
  {1958})}\BibitemShut {NoStop}%
\bibitem [{\citenamefont {Abrahams}\ \emph {et~al.}(1979)\citenamefont
  {Abrahams}, \citenamefont {Anderson}, \citenamefont {Licciardello},\ and\
  \citenamefont {Ramakrishnan}}]{abrahams_scaling_1979}%
  \BibitemOpen
  \bibfield  {author} {\bibinfo {author} {\bibfnamefont {E.}~\bibnamefont
  {Abrahams}}, \bibinfo {author} {\bibfnamefont {P.~W.}\ \bibnamefont
  {Anderson}}, \bibinfo {author} {\bibfnamefont {D.~C.}\ \bibnamefont
  {Licciardello}},\ and\ \bibinfo {author} {\bibfnamefont {T.~V.}\ \bibnamefont
  {Ramakrishnan}},\ }\bibfield  {title} {\bibinfo {title} {Scaling theory of
  localization: {Absence} of quantum diffusion in two dimensions},\ }\href
  {https://doi.org/10.1103/PhysRevLett.42.673} {\bibfield  {journal} {\bibinfo
  {journal} {Phys. Rev. Lett.}\ }\textbf {\bibinfo {volume} {42}},\ \bibinfo
  {pages} {673} (\bibinfo {year} {1979})}\BibitemShut {NoStop}%
\bibitem [{\citenamefont {Lee}\ and\ \citenamefont
  {Ramakrishnan}(1985)}]{lee1985disordered}%
  \BibitemOpen
  \bibfield  {author} {\bibinfo {author} {\bibfnamefont {P.~A.}\ \bibnamefont
  {Lee}}\ and\ \bibinfo {author} {\bibfnamefont {T.~V.}\ \bibnamefont
  {Ramakrishnan}},\ }\bibfield  {title} {\bibinfo {title} {Disordered
  electronic systems},\ }\href {https://doi.org/10.1103/RevModPhys.57.287}
  {\bibfield  {journal} {\bibinfo  {journal} {Rev. Mod. Phys.}\ }\textbf
  {\bibinfo {volume} {57}},\ \bibinfo {pages} {287} (\bibinfo {year}
  {1985})}\BibitemShut {NoStop}%
\bibitem [{\citenamefont {Evers}\ and\ \citenamefont
  {Mirlin}(2008)}]{evers2008anderson}%
  \BibitemOpen
  \bibfield  {author} {\bibinfo {author} {\bibfnamefont {F.}~\bibnamefont
  {Evers}}\ and\ \bibinfo {author} {\bibfnamefont {A.~D.}\ \bibnamefont
  {Mirlin}},\ }\bibfield  {title} {\bibinfo {title} {Anderson transitions},\
  }\href {https://doi.org/10.1103/RevModPhys.80.1355} {\bibfield  {journal}
  {\bibinfo  {journal} {Rev. Mod. Phys.}\ }\textbf {\bibinfo {volume} {80}},\
  \bibinfo {pages} {1355} (\bibinfo {year} {2008})}\BibitemShut {NoStop}%
\bibitem [{\citenamefont {Yamilov}\ \emph {et~al.}(2023)\citenamefont
  {Yamilov}, \citenamefont {Skipetrov}, \citenamefont {Hughes}, \citenamefont
  {Minkov}, \citenamefont {Yu},\ and\ \citenamefont
  {Cao}}]{yamilov2023anderson}%
  \BibitemOpen
  \bibfield  {author} {\bibinfo {author} {\bibfnamefont {A.}~\bibnamefont
  {Yamilov}}, \bibinfo {author} {\bibfnamefont {S.~E.}\ \bibnamefont
  {Skipetrov}}, \bibinfo {author} {\bibfnamefont {T.~W.}\ \bibnamefont
  {Hughes}}, \bibinfo {author} {\bibfnamefont {M.}~\bibnamefont {Minkov}},
  \bibinfo {author} {\bibfnamefont {Z.}~\bibnamefont {Yu}},\ and\ \bibinfo
  {author} {\bibfnamefont {H.}~\bibnamefont {Cao}},\ }\bibfield  {title}
  {\bibinfo {title} {Anderson localization of electromagnetic waves in three
  dimensions},\ }\href
  {https://doi.org/https://doi.org/10.1038/s41567-023-02091-7} {\bibfield
  {journal} {\bibinfo  {journal} {Nat. Phys.}\ }\textbf {\bibinfo {volume}
  {19}},\ \bibinfo {pages} {1308} (\bibinfo {year} {2023})}\BibitemShut
  {NoStop}%
\bibitem [{\citenamefont {Li}\ \emph {et~al.}(2009)\citenamefont {Li},
  \citenamefont {Chu}, \citenamefont {Jain},\ and\ \citenamefont
  {Shen}}]{li_topological_2009}%
  \BibitemOpen
  \bibfield  {author} {\bibinfo {author} {\bibfnamefont {J.}~\bibnamefont
  {Li}}, \bibinfo {author} {\bibfnamefont {R.-L.}\ \bibnamefont {Chu}},
  \bibinfo {author} {\bibfnamefont {J.~K.}\ \bibnamefont {Jain}},\ and\
  \bibinfo {author} {\bibfnamefont {S.-Q.}\ \bibnamefont {Shen}},\ }\bibfield
  {title} {\bibinfo {title} {Topological {Anderson} insulator},\ }\href
  {https://doi.org/10.1103/PhysRevLett.102.136806} {\bibfield  {journal}
  {\bibinfo  {journal} {Phys. Rev. Lett.}\ }\textbf {\bibinfo {volume} {102}},\
  \bibinfo {pages} {136806} (\bibinfo {year} {2009})}\BibitemShut {NoStop}%
\bibitem [{\citenamefont {Groth}\ \emph {et~al.}(2009)\citenamefont {Groth},
  \citenamefont {Wimmer}, \citenamefont {Akhmerov}, \citenamefont
  {Tworzyd{\l}o},\ and\ \citenamefont {Beenakker}}]{groth_theory_2009}%
  \BibitemOpen
  \bibfield  {author} {\bibinfo {author} {\bibfnamefont {C.~W.}\ \bibnamefont
  {Groth}}, \bibinfo {author} {\bibfnamefont {M.}~\bibnamefont {Wimmer}},
  \bibinfo {author} {\bibfnamefont {A.~R.}\ \bibnamefont {Akhmerov}}, \bibinfo
  {author} {\bibfnamefont {J.}~\bibnamefont {Tworzyd{\l}o}},\ and\ \bibinfo
  {author} {\bibfnamefont {C.~W.~J.}\ \bibnamefont {Beenakker}},\ }\bibfield
  {title} {\bibinfo {title} {Theory of the topological {Anderson} insulator},\
  }\href {https://doi.org/10.1103/PhysRevLett.103.196805} {\bibfield  {journal}
  {\bibinfo  {journal} {Phys. Rev. Lett.}\ }\textbf {\bibinfo {volume} {103}},\
  \bibinfo {pages} {196805} (\bibinfo {year} {2009})}\BibitemShut {NoStop}%
\bibitem [{\citenamefont {Jiang}\ \emph {et~al.}(2009)\citenamefont {Jiang},
  \citenamefont {Wang}, \citenamefont {Sun},\ and\ \citenamefont
  {Xie}}]{jiang2009numerical}%
  \BibitemOpen
  \bibfield  {author} {\bibinfo {author} {\bibfnamefont {H.}~\bibnamefont
  {Jiang}}, \bibinfo {author} {\bibfnamefont {L.}~\bibnamefont {Wang}},
  \bibinfo {author} {\bibfnamefont {Q.-f.}\ \bibnamefont {Sun}},\ and\ \bibinfo
  {author} {\bibfnamefont {X.}~\bibnamefont {Xie}},\ }\bibfield  {title}
  {\bibinfo {title} {Numerical study of the topological {A}nderson insulator in
  {H}g{T}e/{C}d{T}e quantum wells},\ }\href
  {https://doi.org/10.1103/PhysRevB.80.165316} {\bibfield  {journal} {\bibinfo
  {journal} {Phys. Rev. B}\ }\textbf {\bibinfo {volume} {80}},\ \bibinfo
  {pages} {165316} (\bibinfo {year} {2009})}\BibitemShut {NoStop}%
\bibitem [{\citenamefont {Meier}\ \emph {et~al.}(2018)\citenamefont {Meier},
  \citenamefont {An}, \citenamefont {Dauphin}, \citenamefont {Maffei},
  \citenamefont {Massignan}, \citenamefont {Hughes},\ and\ \citenamefont
  {Gadway}}]{meier_observation_2018}%
  \BibitemOpen
  \bibfield  {author} {\bibinfo {author} {\bibfnamefont {E.~J.}\ \bibnamefont
  {Meier}}, \bibinfo {author} {\bibfnamefont {F.~A.}\ \bibnamefont {An}},
  \bibinfo {author} {\bibfnamefont {A.}~\bibnamefont {Dauphin}}, \bibinfo
  {author} {\bibfnamefont {M.}~\bibnamefont {Maffei}}, \bibinfo {author}
  {\bibfnamefont {P.}~\bibnamefont {Massignan}}, \bibinfo {author}
  {\bibfnamefont {T.~L.}\ \bibnamefont {Hughes}},\ and\ \bibinfo {author}
  {\bibfnamefont {B.}~\bibnamefont {Gadway}},\ }\bibfield  {title} {\bibinfo
  {title} {Observation of the topological {Anderson} insulator in disordered
  atomic wires},\ }\href {https://doi.org/10.1126/science.aat3406} {\bibfield
  {journal} {\bibinfo  {journal} {Science}\ }\textbf {\bibinfo {volume}
  {362}},\ \bibinfo {pages} {929} (\bibinfo {year} {2018})}\BibitemShut
  {NoStop}%
\bibitem [{\citenamefont {St{\"u}tzer}\ \emph {et~al.}(2018)\citenamefont
  {St{\"u}tzer}, \citenamefont {Plotnik}, \citenamefont {Lumer}, \citenamefont
  {Titum}, \citenamefont {Lindner}, \citenamefont {Segev}, \citenamefont
  {Rechtsman},\ and\ \citenamefont {Szameit}}]{stutzer_photonic_2018}%
  \BibitemOpen
  \bibfield  {author} {\bibinfo {author} {\bibfnamefont {S.}~\bibnamefont
  {St{\"u}tzer}}, \bibinfo {author} {\bibfnamefont {Y.}~\bibnamefont
  {Plotnik}}, \bibinfo {author} {\bibfnamefont {Y.}~\bibnamefont {Lumer}},
  \bibinfo {author} {\bibfnamefont {P.}~\bibnamefont {Titum}}, \bibinfo
  {author} {\bibfnamefont {N.~H.}\ \bibnamefont {Lindner}}, \bibinfo {author}
  {\bibfnamefont {M.}~\bibnamefont {Segev}}, \bibinfo {author} {\bibfnamefont
  {M.~C.}\ \bibnamefont {Rechtsman}},\ and\ \bibinfo {author} {\bibfnamefont
  {A.}~\bibnamefont {Szameit}},\ }\bibfield  {title} {\bibinfo {title}
  {Photonic topological {{Anderson}} insulators},\ }\href
  {https://doi.org/10.1038/s41586-018-0418-2} {\bibfield  {journal} {\bibinfo
  {journal} {Nature}\ }\textbf {\bibinfo {volume} {560}},\ \bibinfo {pages}
  {461} (\bibinfo {year} {2018})}\BibitemShut {NoStop}%
\bibitem [{\citenamefont {Hasan}\ and\ \citenamefont
  {Kane}(2010)}]{hasan_colloquium_2010}%
  \BibitemOpen
  \bibfield  {author} {\bibinfo {author} {\bibfnamefont {M.~Z.}\ \bibnamefont
  {Hasan}}\ and\ \bibinfo {author} {\bibfnamefont {C.~L.}\ \bibnamefont
  {Kane}},\ }\bibfield  {title} {\bibinfo {title} {{\emph{Colloquium}} :
  {{Topological}} insulators},\ }\href
  {https://doi.org/10.1103/RevModPhys.82.3045} {\bibfield  {journal} {\bibinfo
  {journal} {Rev. Mod. Phys.}\ }\textbf {\bibinfo {volume} {82}},\ \bibinfo
  {pages} {3045} (\bibinfo {year} {2010})}\BibitemShut {NoStop}%
\bibitem [{\citenamefont {Qi}\ and\ \citenamefont
  {Zhang}(2011)}]{qi_topological_2011}%
  \BibitemOpen
  \bibfield  {author} {\bibinfo {author} {\bibfnamefont {X.-L.}\ \bibnamefont
  {Qi}}\ and\ \bibinfo {author} {\bibfnamefont {S.-C.}\ \bibnamefont {Zhang}},\
  }\bibfield  {title} {\bibinfo {title} {Topological insulators and
  superconductors},\ }\href {https://doi.org/10.1103/RevModPhys.83.1057}
  {\bibfield  {journal} {\bibinfo  {journal} {Rev. Mod. Phys.}\ }\textbf
  {\bibinfo {volume} {83}},\ \bibinfo {pages} {1057} (\bibinfo {year}
  {2011})}\BibitemShut {NoStop}%
\bibitem [{\citenamefont {Guo}\ \emph {et~al.}(2010)\citenamefont {Guo},
  \citenamefont {Rosenberg}, \citenamefont {Refael},\ and\ \citenamefont
  {Franz}}]{guo_topological_2010}%
  \BibitemOpen
  \bibfield  {author} {\bibinfo {author} {\bibfnamefont {H.-M.}\ \bibnamefont
  {Guo}}, \bibinfo {author} {\bibfnamefont {G.}~\bibnamefont {Rosenberg}},
  \bibinfo {author} {\bibfnamefont {G.}~\bibnamefont {Refael}},\ and\ \bibinfo
  {author} {\bibfnamefont {M.}~\bibnamefont {Franz}},\ }\bibfield  {title}
  {\bibinfo {title} {Topological {Anderson} insulator in three dimensions},\
  }\href {https://doi.org/10.1103/PhysRevLett.105.216601} {\bibfield  {journal}
  {\bibinfo  {journal} {Phys. Rev. Lett.}\ }\textbf {\bibinfo {volume} {105}},\
  \bibinfo {pages} {216601} (\bibinfo {year} {2010})}\BibitemShut {NoStop}%
\bibitem [{\citenamefont {{Mondragon-Shem}}\ \emph {et~al.}(2014)\citenamefont
  {{Mondragon-Shem}}, \citenamefont {Hughes}, \citenamefont {Song},\ and\
  \citenamefont {Prodan}}]{mondragon-shem_topological_2014}%
  \BibitemOpen
  \bibfield  {author} {\bibinfo {author} {\bibfnamefont {I.}~\bibnamefont
  {{Mondragon-Shem}}}, \bibinfo {author} {\bibfnamefont {T.~L.}\ \bibnamefont
  {Hughes}}, \bibinfo {author} {\bibfnamefont {J.}~\bibnamefont {Song}},\ and\
  \bibinfo {author} {\bibfnamefont {E.}~\bibnamefont {Prodan}},\ }\bibfield
  {title} {\bibinfo {title} {Topological criticality in the chiral-symmetric
  {{AIII}} class at strong disorder},\ }\href
  {https://doi.org/10.1103/PhysRevLett.113.046802} {\bibfield  {journal}
  {\bibinfo  {journal} {Phys. Rev. Lett.}\ }\textbf {\bibinfo {volume} {113}},\
  \bibinfo {pages} {046802} (\bibinfo {year} {2014})}\BibitemShut {NoStop}%
\bibitem [{\citenamefont {Titum}\ \emph {et~al.}(2015)\citenamefont {Titum},
  \citenamefont {Lindner}, \citenamefont {Rechtsman},\ and\ \citenamefont
  {Refael}}]{titum_disorderinduced_2015}%
  \BibitemOpen
  \bibfield  {author} {\bibinfo {author} {\bibfnamefont {P.}~\bibnamefont
  {Titum}}, \bibinfo {author} {\bibfnamefont {N.~H.}\ \bibnamefont {Lindner}},
  \bibinfo {author} {\bibfnamefont {M.~C.}\ \bibnamefont {Rechtsman}},\ and\
  \bibinfo {author} {\bibfnamefont {G.}~\bibnamefont {Refael}},\ }\bibfield
  {title} {\bibinfo {title} {Disorder-induced {Floquet} topological
  insulators},\ }\href {https://doi.org/10.1103/PhysRevLett.114.056801}
  {\bibfield  {journal} {\bibinfo  {journal} {Phys. Rev. Lett.}\ }\textbf
  {\bibinfo {volume} {114}},\ \bibinfo {pages} {056801} (\bibinfo {year}
  {2015})}\BibitemShut {NoStop}%
\bibitem [{\citenamefont {Su}\ \emph {et~al.}(2016)\citenamefont {Su},
  \citenamefont {Avishai},\ and\ \citenamefont {Wang}}]{su2016topological}%
  \BibitemOpen
  \bibfield  {author} {\bibinfo {author} {\bibfnamefont {Y.}~\bibnamefont
  {Su}}, \bibinfo {author} {\bibfnamefont {Y.}~\bibnamefont {Avishai}},\ and\
  \bibinfo {author} {\bibfnamefont {X.}~\bibnamefont {Wang}},\ }\bibfield
  {title} {\bibinfo {title} {Topological {A}nderson insulators in systems
  without time-reversal symmetry},\ }\href
  {https://doi.org/10.1103/PhysRevB.93.214206} {\bibfield  {journal} {\bibinfo
  {journal} {Phys. Rev. B}\ }\textbf {\bibinfo {volume} {93}},\ \bibinfo
  {pages} {214206} (\bibinfo {year} {2016})}\BibitemShut {NoStop}%
\bibitem [{\citenamefont {Liu}\ \emph {et~al.}(2020)\citenamefont {Liu},
  \citenamefont {Yang}, \citenamefont {Ren}, \citenamefont {Xue}, \citenamefont
  {Lin}, \citenamefont {Hu}, \citenamefont {Sun}, \citenamefont {Peng},
  \citenamefont {Zhou}, \citenamefont {Chong},\ and\ \citenamefont
  {Zhang}}]{liu_topological_2020}%
  \BibitemOpen
  \bibfield  {author} {\bibinfo {author} {\bibfnamefont {G.-G.}\ \bibnamefont
  {Liu}}, \bibinfo {author} {\bibfnamefont {Y.}~\bibnamefont {Yang}}, \bibinfo
  {author} {\bibfnamefont {X.}~\bibnamefont {Ren}}, \bibinfo {author}
  {\bibfnamefont {H.}~\bibnamefont {Xue}}, \bibinfo {author} {\bibfnamefont
  {X.}~\bibnamefont {Lin}}, \bibinfo {author} {\bibfnamefont {Y.-H.}\
  \bibnamefont {Hu}}, \bibinfo {author} {\bibfnamefont {H.-x.}\ \bibnamefont
  {Sun}}, \bibinfo {author} {\bibfnamefont {B.}~\bibnamefont {Peng}}, \bibinfo
  {author} {\bibfnamefont {P.}~\bibnamefont {Zhou}}, \bibinfo {author}
  {\bibfnamefont {Y.}~\bibnamefont {Chong}},\ and\ \bibinfo {author}
  {\bibfnamefont {B.}~\bibnamefont {Zhang}},\ }\bibfield  {title} {\bibinfo
  {title} {Topological {Anderson} insulator in disordered photonic crystals},\
  }\href {https://doi.org/10.1103/PhysRevLett.125.133603} {\bibfield  {journal}
  {\bibinfo  {journal} {Phys. Rev. Lett.}\ }\textbf {\bibinfo {volume} {125}},\
  \bibinfo {pages} {133603} (\bibinfo {year} {2020})}\BibitemShut {NoStop}%
\bibitem [{\citenamefont {Zangeneh-Nejad}\ and\ \citenamefont
  {Fleury}(2020)}]{zangeneh2020disorder}%
  \BibitemOpen
  \bibfield  {author} {\bibinfo {author} {\bibfnamefont {F.}~\bibnamefont
  {Zangeneh-Nejad}}\ and\ \bibinfo {author} {\bibfnamefont {R.}~\bibnamefont
  {Fleury}},\ }\bibfield  {title} {\bibinfo {title} {Disorder-induced signal
  filtering with topological metamaterials},\ }\href
  {https://doi.org/10.1002/adma.202001034} {\bibfield  {journal} {\bibinfo
  {journal} {Adv. Mater.}\ }\textbf {\bibinfo {volume} {32}},\ \bibinfo {pages}
  {2001034} (\bibinfo {year} {2020})}\BibitemShut {NoStop}%
\bibitem [{\citenamefont {Zhang}\ \emph
  {et~al.}(2021{\natexlab{a}})\citenamefont {Zhang}, \citenamefont {Zou},
  \citenamefont {Pei}, \citenamefont {He}, \citenamefont {Bao}, \citenamefont
  {Sun},\ and\ \citenamefont {Zhang}}]{zhang_experimental_2021}%
  \BibitemOpen
  \bibfield  {author} {\bibinfo {author} {\bibfnamefont {W.}~\bibnamefont
  {Zhang}}, \bibinfo {author} {\bibfnamefont {D.}~\bibnamefont {Zou}}, \bibinfo
  {author} {\bibfnamefont {Q.}~\bibnamefont {Pei}}, \bibinfo {author}
  {\bibfnamefont {W.}~\bibnamefont {He}}, \bibinfo {author} {\bibfnamefont
  {J.}~\bibnamefont {Bao}}, \bibinfo {author} {\bibfnamefont {H.}~\bibnamefont
  {Sun}},\ and\ \bibinfo {author} {\bibfnamefont {X.}~\bibnamefont {Zhang}},\
  }\bibfield  {title} {\bibinfo {title} {Experimental observation of
  higher-order topological {Anderson} insulators},\ }\href
  {https://doi.org/10.1103/PhysRevLett.126.146802} {\bibfield  {journal}
  {\bibinfo  {journal} {Phys. Rev. Lett.}\ }\textbf {\bibinfo {volume} {126}},\
  \bibinfo {pages} {146802} (\bibinfo {year} {2021}{\natexlab{a}})}\BibitemShut
  {NoStop}%
\bibitem [{\citenamefont {Cui}\ \emph {et~al.}(2022)\citenamefont {Cui},
  \citenamefont {Zhang}, \citenamefont {Zhang},\ and\ \citenamefont
  {Chan}}]{cui_photonic_2022}%
  \BibitemOpen
  \bibfield  {author} {\bibinfo {author} {\bibfnamefont {X.}~\bibnamefont
  {Cui}}, \bibinfo {author} {\bibfnamefont {R.-Y.}\ \bibnamefont {Zhang}},
  \bibinfo {author} {\bibfnamefont {Z.-Q.}\ \bibnamefont {Zhang}},\ and\
  \bibinfo {author} {\bibfnamefont {C.~T.}\ \bibnamefont {Chan}},\ }\bibfield
  {title} {\bibinfo {title} {Photonic ${Z}_2$ topological {A}nderson
  insulators},\ }\href {https://doi.org/10.1103/PhysRevLett.129.043902}
  {\bibfield  {journal} {\bibinfo  {journal} {Phys. Rev. Lett.}\ }\textbf
  {\bibinfo {volume} {129}},\ \bibinfo {pages} {043902} (\bibinfo {year}
  {2022})}\BibitemShut {NoStop}%
\bibitem [{\citenamefont {Gao}\ \emph {et~al.}(2022)\citenamefont {Gao},
  \citenamefont {Xu}, \citenamefont {Smirnova}, \citenamefont {Leykam},
  \citenamefont {Gyger}, \citenamefont {Zhou}, \citenamefont {Steinhauer},
  \citenamefont {Zwiller},\ and\ \citenamefont
  {Elshaari}}]{gao_observation_2022}%
  \BibitemOpen
  \bibfield  {author} {\bibinfo {author} {\bibfnamefont {J.}~\bibnamefont
  {Gao}}, \bibinfo {author} {\bibfnamefont {Z.-S.}\ \bibnamefont {Xu}},
  \bibinfo {author} {\bibfnamefont {D.~A.}\ \bibnamefont {Smirnova}}, \bibinfo
  {author} {\bibfnamefont {D.}~\bibnamefont {Leykam}}, \bibinfo {author}
  {\bibfnamefont {S.}~\bibnamefont {Gyger}}, \bibinfo {author} {\bibfnamefont
  {W.-H.}\ \bibnamefont {Zhou}}, \bibinfo {author} {\bibfnamefont
  {S.}~\bibnamefont {Steinhauer}}, \bibinfo {author} {\bibfnamefont
  {V.}~\bibnamefont {Zwiller}},\ and\ \bibinfo {author} {\bibfnamefont {A.~W.}\
  \bibnamefont {Elshaari}},\ }\bibfield  {title} {\bibinfo {title} {Observation
  of {{Anderson}} phase in a topological photonic circuit},\ }\href
  {https://doi.org/10.1103/PhysRevResearch.4.033222} {\bibfield  {journal}
  {\bibinfo  {journal} {Phys. Rev. Res.}\ }\textbf {\bibinfo {volume} {4}},\
  \bibinfo {pages} {033222} (\bibinfo {year} {2022})}\BibitemShut {NoStop}%
\bibitem [{\citenamefont {Chen}\ \emph {et~al.}(2023)\citenamefont {Chen},
  \citenamefont {Yi},\ and\ \citenamefont {Zhou}}]{chen2023four}%
  \BibitemOpen
  \bibfield  {author} {\bibinfo {author} {\bibfnamefont {R.}~\bibnamefont
  {Chen}}, \bibinfo {author} {\bibfnamefont {X.-X.}\ \bibnamefont {Yi}},\ and\
  \bibinfo {author} {\bibfnamefont {B.}~\bibnamefont {Zhou}},\ }\bibfield
  {title} {\bibinfo {title} {Four-dimensional topological {A}nderson insulator
  with an emergent second {C}hern number},\ }\href
  {https://doi.org/10.1103/PhysRevB.108.085306} {\bibfield  {journal} {\bibinfo
   {journal} {Phys. Rev. B}\ }\textbf {\bibinfo {volume} {108}},\ \bibinfo
  {pages} {085306} (\bibinfo {year} {2023})}\BibitemShut {NoStop}%
\bibitem [{\citenamefont {Tang}\ \emph {et~al.}(2020)\citenamefont {Tang},
  \citenamefont {Zhang}, \citenamefont {Zhang},\ and\ \citenamefont
  {Zhang}}]{tang_topological_2020}%
  \BibitemOpen
  \bibfield  {author} {\bibinfo {author} {\bibfnamefont {L.-Z.}\ \bibnamefont
  {Tang}}, \bibinfo {author} {\bibfnamefont {L.-F.}\ \bibnamefont {Zhang}},
  \bibinfo {author} {\bibfnamefont {G.-Q.}\ \bibnamefont {Zhang}},\ and\
  \bibinfo {author} {\bibfnamefont {D.-W.}\ \bibnamefont {Zhang}},\ }\bibfield
  {title} {\bibinfo {title} {Topological {{Anderson}} insulators in
  two-dimensional non-{{Hermitian}} disordered systems},\ }\href
  {https://doi.org/10.1103/PhysRevA.101.063612} {\bibfield  {journal} {\bibinfo
   {journal} {Phys. Rev. A}\ }\textbf {\bibinfo {volume} {101}},\ \bibinfo
  {pages} {063612} (\bibinfo {year} {2020})}\BibitemShut {NoStop}%
\bibitem [{\citenamefont {Zhang}\ \emph
  {et~al.}(2020{\natexlab{a}})\citenamefont {Zhang}, \citenamefont {Tang},
  \citenamefont {Lang}, \citenamefont {Yan},\ and\ \citenamefont
  {Zhu}}]{zhang_nonhermitian_2020}%
  \BibitemOpen
  \bibfield  {author} {\bibinfo {author} {\bibfnamefont {D.-W.}\ \bibnamefont
  {Zhang}}, \bibinfo {author} {\bibfnamefont {L.-Z.}\ \bibnamefont {Tang}},
  \bibinfo {author} {\bibfnamefont {L.-J.}\ \bibnamefont {Lang}}, \bibinfo
  {author} {\bibfnamefont {H.}~\bibnamefont {Yan}},\ and\ \bibinfo {author}
  {\bibfnamefont {S.-L.}\ \bibnamefont {Zhu}},\ }\bibfield  {title} {\bibinfo
  {title} {Non-{{Hermitian}} topological {{Anderson}} insulators},\ }\href
  {https://doi.org/10.1007/s11433-020-1521-9} {\bibfield  {journal} {\bibinfo
  {journal} {Sci. China Phys. Mech. Astron.}\ }\textbf {\bibinfo {volume}
  {63}},\ \bibinfo {pages} {267062} (\bibinfo {year}
  {2020}{\natexlab{a}})}\BibitemShut {NoStop}%
\bibitem [{\citenamefont {Liu}\ \emph {et~al.}(2021)\citenamefont {Liu},
  \citenamefont {Zhou}, \citenamefont {Wu}, \citenamefont {Zhang},\ and\
  \citenamefont {Jiang}}]{liu_realspace_2021}%
  \BibitemOpen
  \bibfield  {author} {\bibinfo {author} {\bibfnamefont {H.}~\bibnamefont
  {Liu}}, \bibinfo {author} {\bibfnamefont {J.-K.}\ \bibnamefont {Zhou}},
  \bibinfo {author} {\bibfnamefont {B.-L.}\ \bibnamefont {Wu}}, \bibinfo
  {author} {\bibfnamefont {Z.-Q.}\ \bibnamefont {Zhang}},\ and\ \bibinfo
  {author} {\bibfnamefont {H.}~\bibnamefont {Jiang}},\ }\bibfield  {title}
  {\bibinfo {title} {Real-space topological invariant and higher-order
  topological {{Anderson}} insulator in two-dimensional non-{{Hermitian}}
  systems},\ }\href {https://doi.org/10.1103/PhysRevB.103.224203} {\bibfield
  {journal} {\bibinfo  {journal} {Phys. Rev. B}\ }\textbf {\bibinfo {volume}
  {103}},\ \bibinfo {pages} {224203} (\bibinfo {year} {2021})}\BibitemShut
  {NoStop}%
\bibitem [{\citenamefont {Zhang}\ \emph
  {et~al.}(2021{\natexlab{b}})\citenamefont {Zhang}, \citenamefont {Sheng},\
  and\ \citenamefont {Xing}}]{zhang_nonhermitian_2021}%
  \BibitemOpen
  \bibfield  {author} {\bibinfo {author} {\bibfnamefont {C.}~\bibnamefont
  {Zhang}}, \bibinfo {author} {\bibfnamefont {L.}~\bibnamefont {Sheng}},\ and\
  \bibinfo {author} {\bibfnamefont {D.}~\bibnamefont {Xing}},\ }\bibfield
  {title} {\bibinfo {title} {Non-{{Hermitian}} disorder-driven topological
  transition in a dimerized {{Kitaev}} superconductor chain},\ }\href
  {https://doi.org/10.1103/PhysRevB.103.224207} {\bibfield  {journal} {\bibinfo
   {journal} {Phys. Rev. B}\ }\textbf {\bibinfo {volume} {103}},\ \bibinfo
  {pages} {224207} (\bibinfo {year} {2021}{\natexlab{b}})}\BibitemShut
  {NoStop}%
\bibitem [{\citenamefont {Luo}\ and\ \citenamefont
  {Zhang}(2023)}]{luo_photonic_2022}%
  \BibitemOpen
  \bibfield  {author} {\bibinfo {author} {\bibfnamefont {X.-W.}\ \bibnamefont
  {Luo}}\ and\ \bibinfo {author} {\bibfnamefont {C.}~\bibnamefont {Zhang}},\
  }\bibfield  {title} {\bibinfo {title} {Photonic topological insulators
  induced by {non-Hermitian} disorders in a coupled-cavity array},\ }\bibfield
  {journal} {\bibinfo  {journal} {Appl. Phys. Lett.}\ }\textbf {\bibinfo
  {volume} {123}},\ \href {https://doi.org/https://doi.org/10.1063/5.0153523}
  {https://doi.org/10.1063/5.0153523} (\bibinfo {year} {2023})\BibitemShut
  {NoStop}%
\bibitem [{\citenamefont {Lin}\ \emph {et~al.}(2022)\citenamefont {Lin},
  \citenamefont {Li}, \citenamefont {Xiao}, \citenamefont {Wang}, \citenamefont
  {Yi},\ and\ \citenamefont {Xue}}]{lin_observation_2022}%
  \BibitemOpen
  \bibfield  {author} {\bibinfo {author} {\bibfnamefont {Q.}~\bibnamefont
  {Lin}}, \bibinfo {author} {\bibfnamefont {T.}~\bibnamefont {Li}}, \bibinfo
  {author} {\bibfnamefont {L.}~\bibnamefont {Xiao}}, \bibinfo {author}
  {\bibfnamefont {K.}~\bibnamefont {Wang}}, \bibinfo {author} {\bibfnamefont
  {W.}~\bibnamefont {Yi}},\ and\ \bibinfo {author} {\bibfnamefont
  {P.}~\bibnamefont {Xue}},\ }\bibfield  {title} {\bibinfo {title} {Observation
  of non-{{Hermitian}} topological {{Anderson}} insulator in quantum
  dynamics},\ }\href {https://doi.org/10.1038/s41467-022-30938-9} {\bibfield
  {journal} {\bibinfo  {journal} {Nat. Commun.}\ }\textbf {\bibinfo {volume}
  {13}},\ \bibinfo {pages} {3229} (\bibinfo {year} {2022})}\BibitemShut
  {NoStop}%
\bibitem [{\citenamefont {Gu}\ \emph {et~al.}(2023)\citenamefont {Gu},
  \citenamefont {Gao}, \citenamefont {Xue}, \citenamefont {Wang}, \citenamefont
  {Guo}, \citenamefont {Su}, \citenamefont {Zhang},\ and\ \citenamefont
  {Zhu}}]{gu_observation_2023}%
  \BibitemOpen
  \bibfield  {author} {\bibinfo {author} {\bibfnamefont {Z.}~\bibnamefont
  {Gu}}, \bibinfo {author} {\bibfnamefont {H.}~\bibnamefont {Gao}}, \bibinfo
  {author} {\bibfnamefont {H.}~\bibnamefont {Xue}}, \bibinfo {author}
  {\bibfnamefont {D.}~\bibnamefont {Wang}}, \bibinfo {author} {\bibfnamefont
  {J.}~\bibnamefont {Guo}}, \bibinfo {author} {\bibfnamefont {Z.}~\bibnamefont
  {Su}}, \bibinfo {author} {\bibfnamefont {B.}~\bibnamefont {Zhang}},\ and\
  \bibinfo {author} {\bibfnamefont {J.}~\bibnamefont {Zhu}},\ }\bibfield
  {title} {\bibinfo {title} {Observation of an acoustic non-{{Hermitian}}
  topological {{Anderson}} insulator},\ }\href
  {https://doi.org/10.1007/s11433-023-2159-4} {\bibfield  {journal} {\bibinfo
  {journal} {Sci. China Phys. Mech. Astron.}\ }\textbf {\bibinfo {volume}
  {66}},\ \bibinfo {pages} {294311} (\bibinfo {year} {2023})}\BibitemShut
  {NoStop}%
\bibitem [{\citenamefont {Mo}\ \emph {et~al.}(2022)\citenamefont {Mo},
  \citenamefont {Sun}, \citenamefont {Li}, \citenamefont {Ruan},\ and\
  \citenamefont {Yang}}]{mo_imaginary-disorder-induced_2022}%
  \BibitemOpen
  \bibfield  {author} {\bibinfo {author} {\bibfnamefont {Q.}~\bibnamefont
  {Mo}}, \bibinfo {author} {\bibfnamefont {Y.}~\bibnamefont {Sun}}, \bibinfo
  {author} {\bibfnamefont {J.}~\bibnamefont {Li}}, \bibinfo {author}
  {\bibfnamefont {Z.}~\bibnamefont {Ruan}},\ and\ \bibinfo {author}
  {\bibfnamefont {Z.}~\bibnamefont {Yang}},\ }\bibfield  {title} {\bibinfo
  {title} {Imaginary-disorder-induced topological phase transitions},\ }\href
  {https://doi.org/10.1103/PhysRevApplied.18.064079} {\bibfield  {journal}
  {\bibinfo  {journal} {Phys. Rev. Applied}\ }\textbf {\bibinfo {volume}
  {18}},\ \bibinfo {pages} {064079} (\bibinfo {year} {2022})}\BibitemShut
  {NoStop}%
\bibitem [{\citenamefont {Gong}\ \emph {et~al.}(2018)\citenamefont {Gong},
  \citenamefont {Ashida}, \citenamefont {Kawabata}, \citenamefont {Takasan},
  \citenamefont {Higashikawa},\ and\ \citenamefont
  {Ueda}}]{gong_topological_2018}%
  \BibitemOpen
  \bibfield  {author} {\bibinfo {author} {\bibfnamefont {Z.}~\bibnamefont
  {Gong}}, \bibinfo {author} {\bibfnamefont {Y.}~\bibnamefont {Ashida}},
  \bibinfo {author} {\bibfnamefont {K.}~\bibnamefont {Kawabata}}, \bibinfo
  {author} {\bibfnamefont {K.}~\bibnamefont {Takasan}}, \bibinfo {author}
  {\bibfnamefont {S.}~\bibnamefont {Higashikawa}},\ and\ \bibinfo {author}
  {\bibfnamefont {M.}~\bibnamefont {Ueda}},\ }\bibfield  {title} {\bibinfo
  {title} {Topological phases of non-{Hermitian} systems},\ }\href
  {https://doi.org/10.1103/PhysRevX.8.031079} {\bibfield  {journal} {\bibinfo
  {journal} {Phys. Rev. X}\ }\textbf {\bibinfo {volume} {8}},\ \bibinfo {pages}
  {031079} (\bibinfo {year} {2018})}\BibitemShut {NoStop}%
\bibitem [{\citenamefont {Kawabata}\ \emph {et~al.}(2019)\citenamefont
  {Kawabata}, \citenamefont {Shiozaki}, \citenamefont {Ueda},\ and\
  \citenamefont {Sato}}]{kawabata_symmetry_2019}%
  \BibitemOpen
  \bibfield  {author} {\bibinfo {author} {\bibfnamefont {K.}~\bibnamefont
  {Kawabata}}, \bibinfo {author} {\bibfnamefont {K.}~\bibnamefont {Shiozaki}},
  \bibinfo {author} {\bibfnamefont {M.}~\bibnamefont {Ueda}},\ and\ \bibinfo
  {author} {\bibfnamefont {M.}~\bibnamefont {Sato}},\ }\bibfield  {title}
  {\bibinfo {title} {Symmetry and topology in non-{Hermitian} physics},\ }\href
  {https://doi.org/10.1103/PhysRevX.9.041015} {\bibfield  {journal} {\bibinfo
  {journal} {Phys. Rev. X}\ }\textbf {\bibinfo {volume} {9}},\ \bibinfo {pages}
  {041015} (\bibinfo {year} {2019})}\BibitemShut {NoStop}%
\bibitem [{\citenamefont {Yao}\ and\ \citenamefont
  {Wang}(2018)}]{yao_edge_2018}%
  \BibitemOpen
  \bibfield  {author} {\bibinfo {author} {\bibfnamefont {S.}~\bibnamefont
  {Yao}}\ and\ \bibinfo {author} {\bibfnamefont {Z.}~\bibnamefont {Wang}},\
  }\bibfield  {title} {\bibinfo {title} {Edge states and topological invariants
  of non-{Hermitian} systems},\ }\href
  {https://doi.org/10.1103/PhysRevLett.121.086803} {\bibfield  {journal}
  {\bibinfo  {journal} {Phys. Rev. Lett.}\ }\textbf {\bibinfo {volume} {121}},\
  \bibinfo {pages} {086803} (\bibinfo {year} {2018})}\BibitemShut {NoStop}%
\bibitem [{\citenamefont {Song}\ \emph {et~al.}(2019)\citenamefont {Song},
  \citenamefont {Yao},\ and\ \citenamefont {Wang}}]{song_non-hermitian_2019a}%
  \BibitemOpen
  \bibfield  {author} {\bibinfo {author} {\bibfnamefont {F.}~\bibnamefont
  {Song}}, \bibinfo {author} {\bibfnamefont {S.}~\bibnamefont {Yao}},\ and\
  \bibinfo {author} {\bibfnamefont {Z.}~\bibnamefont {Wang}},\ }\bibfield
  {title} {\bibinfo {title} {Non-{Hermitian} topological invariants in real
  space},\ }\href {https://doi.org/10.1103/PhysRevLett.123.246801} {\bibfield
  {journal} {\bibinfo  {journal} {Phys. Rev. Lett.}\ }\textbf {\bibinfo
  {volume} {123}},\ \bibinfo {pages} {246801} (\bibinfo {year}
  {2019})}\BibitemShut {NoStop}%
\bibitem [{\citenamefont {Borgnia}\ \emph {et~al.}(2020)\citenamefont
  {Borgnia}, \citenamefont {Kruchkov},\ and\ \citenamefont
  {Slager}}]{borgnia_non-hermitian_2020}%
  \BibitemOpen
  \bibfield  {author} {\bibinfo {author} {\bibfnamefont {D.~S.}\ \bibnamefont
  {Borgnia}}, \bibinfo {author} {\bibfnamefont {A.~J.}\ \bibnamefont
  {Kruchkov}},\ and\ \bibinfo {author} {\bibfnamefont {R.-J.}\ \bibnamefont
  {Slager}},\ }\bibfield  {title} {\bibinfo {title} {Non-{{Hermitian}} boundary
  modes and topology},\ }\href {https://doi.org/10.1103/PhysRevLett.124.056802}
  {\bibfield  {journal} {\bibinfo  {journal} {Phys. Rev. Lett.}\ }\textbf
  {\bibinfo {volume} {124}},\ \bibinfo {pages} {056802} (\bibinfo {year}
  {2020})}\BibitemShut {NoStop}%
\bibitem [{\citenamefont {Xiao}\ \emph {et~al.}(2020)\citenamefont {Xiao},
  \citenamefont {Deng}, \citenamefont {Wang}, \citenamefont {Zhu},
  \citenamefont {Wang}, \citenamefont {Yi},\ and\ \citenamefont
  {Xue}}]{xiao2020non}%
  \BibitemOpen
  \bibfield  {author} {\bibinfo {author} {\bibfnamefont {L.}~\bibnamefont
  {Xiao}}, \bibinfo {author} {\bibfnamefont {T.}~\bibnamefont {Deng}}, \bibinfo
  {author} {\bibfnamefont {K.}~\bibnamefont {Wang}}, \bibinfo {author}
  {\bibfnamefont {G.}~\bibnamefont {Zhu}}, \bibinfo {author} {\bibfnamefont
  {Z.}~\bibnamefont {Wang}}, \bibinfo {author} {\bibfnamefont {W.}~\bibnamefont
  {Yi}},\ and\ \bibinfo {author} {\bibfnamefont {P.}~\bibnamefont {Xue}},\
  }\bibfield  {title} {\bibinfo {title} {Non-{Hermitian} bulk--boundary
  correspondence in quantum dynamics},\ }\href
  {https://doi.org/https://doi.org/10.1038/s41567-020-0836-6} {\bibfield
  {journal} {\bibinfo  {journal} {Nat. Phys.}\ }\textbf {\bibinfo {volume}
  {16}},\ \bibinfo {pages} {761} (\bibinfo {year} {2020})}\BibitemShut
  {NoStop}%
\bibitem [{\citenamefont {Helbig}\ \emph {et~al.}(2020)\citenamefont {Helbig},
  \citenamefont {Hofmann}, \citenamefont {Imhof}, \citenamefont {Abdelghany},
  \citenamefont {Kiessling}, \citenamefont {Molenkamp}, \citenamefont {Lee},
  \citenamefont {Szameit}, \citenamefont {Greiter},\ and\ \citenamefont
  {Thomale}}]{helbig2020generalized}%
  \BibitemOpen
  \bibfield  {author} {\bibinfo {author} {\bibfnamefont {T.}~\bibnamefont
  {Helbig}}, \bibinfo {author} {\bibfnamefont {T.}~\bibnamefont {Hofmann}},
  \bibinfo {author} {\bibfnamefont {S.}~\bibnamefont {Imhof}}, \bibinfo
  {author} {\bibfnamefont {M.}~\bibnamefont {Abdelghany}}, \bibinfo {author}
  {\bibfnamefont {T.}~\bibnamefont {Kiessling}}, \bibinfo {author}
  {\bibfnamefont {L.}~\bibnamefont {Molenkamp}}, \bibinfo {author}
  {\bibfnamefont {C.}~\bibnamefont {Lee}}, \bibinfo {author} {\bibfnamefont
  {A.}~\bibnamefont {Szameit}}, \bibinfo {author} {\bibfnamefont
  {M.}~\bibnamefont {Greiter}},\ and\ \bibinfo {author} {\bibfnamefont
  {R.}~\bibnamefont {Thomale}},\ }\bibfield  {title} {\bibinfo {title}
  {Generalized bulk--boundary correspondence in non-{Hermitian} topolectrical
  circuits},\ }\href
  {https://doi.org/https://doi.org/10.1038/s41567-020-0922-9} {\bibfield
  {journal} {\bibinfo  {journal} {Nat. Phys.}\ }\textbf {\bibinfo {volume}
  {16}},\ \bibinfo {pages} {747} (\bibinfo {year} {2020})}\BibitemShut
  {NoStop}%
\bibitem [{\citenamefont {Weidemann}\ \emph {et~al.}(2020)\citenamefont
  {Weidemann}, \citenamefont {Kremer}, \citenamefont {Helbig}, \citenamefont
  {Hofmann}, \citenamefont {Stegmaier}, \citenamefont {Greiter}, \citenamefont
  {Thomale},\ and\ \citenamefont {Szameit}}]{weidemann_topological_2020}%
  \BibitemOpen
  \bibfield  {author} {\bibinfo {author} {\bibfnamefont {S.}~\bibnamefont
  {Weidemann}}, \bibinfo {author} {\bibfnamefont {M.}~\bibnamefont {Kremer}},
  \bibinfo {author} {\bibfnamefont {T.}~\bibnamefont {Helbig}}, \bibinfo
  {author} {\bibfnamefont {T.}~\bibnamefont {Hofmann}}, \bibinfo {author}
  {\bibfnamefont {A.}~\bibnamefont {Stegmaier}}, \bibinfo {author}
  {\bibfnamefont {M.}~\bibnamefont {Greiter}}, \bibinfo {author} {\bibfnamefont
  {R.}~\bibnamefont {Thomale}},\ and\ \bibinfo {author} {\bibfnamefont
  {A.}~\bibnamefont {Szameit}},\ }\bibfield  {title} {\bibinfo {title}
  {Topological funneling of light},\ }\href
  {https://doi.org/10.1126/science.aaz8727} {\bibfield  {journal} {\bibinfo
  {journal} {Science}\ }\textbf {\bibinfo {volume} {368}},\ \bibinfo {pages}
  {311} (\bibinfo {year} {2020})}\BibitemShut {NoStop}%
\bibitem [{\citenamefont {Zhang}\ \emph
  {et~al.}(2021{\natexlab{c}})\citenamefont {Zhang}, \citenamefont {Yang},
  \citenamefont {Ge}, \citenamefont {Guan}, \citenamefont {Chen}, \citenamefont
  {Yan}, \citenamefont {Chen}, \citenamefont {Xi}, \citenamefont {Li},
  \citenamefont {Jia}, \citenamefont {Yuan}, \citenamefont {Sun}, \citenamefont
  {Chen},\ and\ \citenamefont {Zhang}}]{zhang_acoustic_2021}%
  \BibitemOpen
  \bibfield  {author} {\bibinfo {author} {\bibfnamefont {L.}~\bibnamefont
  {Zhang}}, \bibinfo {author} {\bibfnamefont {Y.}~\bibnamefont {Yang}},
  \bibinfo {author} {\bibfnamefont {Y.}~\bibnamefont {Ge}}, \bibinfo {author}
  {\bibfnamefont {Y.-J.}\ \bibnamefont {Guan}}, \bibinfo {author}
  {\bibfnamefont {Q.}~\bibnamefont {Chen}}, \bibinfo {author} {\bibfnamefont
  {Q.}~\bibnamefont {Yan}}, \bibinfo {author} {\bibfnamefont {F.}~\bibnamefont
  {Chen}}, \bibinfo {author} {\bibfnamefont {R.}~\bibnamefont {Xi}}, \bibinfo
  {author} {\bibfnamefont {Y.}~\bibnamefont {Li}}, \bibinfo {author}
  {\bibfnamefont {D.}~\bibnamefont {Jia}}, \bibinfo {author} {\bibfnamefont
  {S.-Q.}\ \bibnamefont {Yuan}}, \bibinfo {author} {\bibfnamefont {H.-X.}\
  \bibnamefont {Sun}}, \bibinfo {author} {\bibfnamefont {H.}~\bibnamefont
  {Chen}},\ and\ \bibinfo {author} {\bibfnamefont {B.}~\bibnamefont {Zhang}},\
  }\bibfield  {title} {\bibinfo {title} {Acoustic non-{{Hermitian}} skin effect
  from twisted winding topology},\ }\href
  {https://doi.org/10.1038/s41467-021-26619-8} {\bibfield  {journal} {\bibinfo
  {journal} {Nat. Commun.}\ }\textbf {\bibinfo {volume} {12}},\ \bibinfo
  {pages} {6297} (\bibinfo {year} {2021}{\natexlab{c}})}\BibitemShut {NoStop}%
\bibitem [{\citenamefont {Zheng}\ \emph {et~al.}(2023)\citenamefont {Zheng},
  \citenamefont {Guo}, \citenamefont {Sun}, \citenamefont {Jiang},
  \citenamefont {Li},\ and\ \citenamefont {Chen}}]{zheng2023topological}%
  \BibitemOpen
  \bibfield  {author} {\bibinfo {author} {\bibfnamefont {J.}~\bibnamefont
  {Zheng}}, \bibinfo {author} {\bibfnamefont {Z.}~\bibnamefont {Guo}}, \bibinfo
  {author} {\bibfnamefont {Y.}~\bibnamefont {Sun}}, \bibinfo {author}
  {\bibfnamefont {H.}~\bibnamefont {Jiang}}, \bibinfo {author} {\bibfnamefont
  {Y.}~\bibnamefont {Li}},\ and\ \bibinfo {author} {\bibfnamefont
  {H.}~\bibnamefont {Chen}},\ }\bibfield  {title} {\bibinfo {title}
  {Topological edge modes in one-dimensional photonic artificial structures.},\
  }\href {https://doi.org/10.2528/PIER22101202} {\bibfield  {journal} {\bibinfo
   {journal} {Prog. Electromagn. Res.}\ }\textbf {\bibinfo {volume} {177}},\
  \bibinfo {pages} {1} (\bibinfo {year} {2023})}\BibitemShut {NoStop}%
\bibitem [{\citenamefont {Lin}\ \emph {et~al.}(2023)\citenamefont {Lin},
  \citenamefont {Tai}, \citenamefont {Li},\ and\ \citenamefont
  {Lee}}]{lin_topological_2023}%
  \BibitemOpen
  \bibfield  {author} {\bibinfo {author} {\bibfnamefont {R.}~\bibnamefont
  {Lin}}, \bibinfo {author} {\bibfnamefont {T.}~\bibnamefont {Tai}}, \bibinfo
  {author} {\bibfnamefont {L.}~\bibnamefont {Li}},\ and\ \bibinfo {author}
  {\bibfnamefont {C.~H.}\ \bibnamefont {Lee}},\ }\bibfield  {title} {\bibinfo
  {title} {Topological non-{{Hermitian}} skin effect},\ }\href
  {https://doi.org/10.1007/s11467-023-1309-z} {\bibfield  {journal} {\bibinfo
  {journal} {Front. Phys.}\ }\textbf {\bibinfo {volume} {18}},\ \bibinfo
  {pages} {53605} (\bibinfo {year} {2023})}\BibitemShut {NoStop}%
\bibitem [{\citenamefont {Claes}\ and\ \citenamefont
  {Hughes}(2021)}]{claes_skin_2021}%
  \BibitemOpen
  \bibfield  {author} {\bibinfo {author} {\bibfnamefont {J.}~\bibnamefont
  {Claes}}\ and\ \bibinfo {author} {\bibfnamefont {T.~L.}\ \bibnamefont
  {Hughes}},\ }\bibfield  {title} {\bibinfo {title} {Skin effect and winding
  number in disordered non-{Hermitian} systems},\ }\href
  {https://doi.org/10.1103/PhysRevB.103.L140201} {\bibfield  {journal}
  {\bibinfo  {journal} {Phys. Rev. B}\ }\textbf {\bibinfo {volume} {103}},\
  \bibinfo {pages} {L140201} (\bibinfo {year} {2021})}\BibitemShut {NoStop}%
\bibitem [{\citenamefont {Kim}\ and\ \citenamefont
  {Park}(2021)}]{kim_disorder-driven_2021}%
  \BibitemOpen
  \bibfield  {author} {\bibinfo {author} {\bibfnamefont {K.-M.}\ \bibnamefont
  {Kim}}\ and\ \bibinfo {author} {\bibfnamefont {M.~J.}\ \bibnamefont {Park}},\
  }\bibfield  {title} {\bibinfo {title} {Disorder-driven phase transition in
  the second-order non-{{Hermitian}} skin effect},\ }\href
  {https://doi.org/10.1103/PhysRevB.104.L121101} {\bibfield  {journal}
  {\bibinfo  {journal} {Phys. Rev. B}\ }\textbf {\bibinfo {volume} {104}},\
  \bibinfo {pages} {L121101} (\bibinfo {year} {2021})}\BibitemShut {NoStop}%
\bibitem [{\citenamefont {Sarkar}\ \emph {et~al.}(2022)\citenamefont {Sarkar},
  \citenamefont {Hegde},\ and\ \citenamefont
  {Narayan}}]{sarkar_interplay_2022}%
  \BibitemOpen
  \bibfield  {author} {\bibinfo {author} {\bibfnamefont {R.}~\bibnamefont
  {Sarkar}}, \bibinfo {author} {\bibfnamefont {S.~S.}\ \bibnamefont {Hegde}},\
  and\ \bibinfo {author} {\bibfnamefont {A.}~\bibnamefont {Narayan}},\
  }\bibfield  {title} {\bibinfo {title} {Interplay of disorder and point-gap
  topology: {Chiral} modes, localization, and non-{{Hermitian Anderson}} skin
  effect in one dimension},\ }\href
  {https://doi.org/10.1103/PhysRevB.106.014207} {\bibfield  {journal} {\bibinfo
   {journal} {Phys. Rev. B}\ }\textbf {\bibinfo {volume} {106}},\ \bibinfo
  {pages} {014207} (\bibinfo {year} {2022})}\BibitemShut {NoStop}%
\bibitem [{\citenamefont {Liu}\ \emph {et~al.}(2023)\citenamefont {Liu},
  \citenamefont {Cai}, \citenamefont {Liu},\ and\ \citenamefont
  {Yang}}]{liu_reentrant_2023}%
  \BibitemOpen
  \bibfield  {author} {\bibinfo {author} {\bibfnamefont {J.}~\bibnamefont
  {Liu}}, \bibinfo {author} {\bibfnamefont {Z.-F.}\ \bibnamefont {Cai}},
  \bibinfo {author} {\bibfnamefont {T.}~\bibnamefont {Liu}},\ and\ \bibinfo
  {author} {\bibfnamefont {Z.}~\bibnamefont {Yang}},\ }\bibfield  {title}
  {\bibinfo {title} {Reentrant non-{{Hermitian}} skin effect in coupled
  non-{{Hermitian}} and {{Hermitian}} chains with correlated disorder},\ }\href
  {https://arxiv.org/abs/2311.03777} {\bibfield  {journal} {\bibinfo  {journal}
  {arXiv: 2311.03777}\ } (\bibinfo {year} {2023})}\BibitemShut {NoStop}%
\bibitem [{\citenamefont {Zhang}\ \emph {et~al.}(2023)\citenamefont {Zhang},
  \citenamefont {Liu}, \citenamefont {Liu}, \citenamefont {Jiang},\ and\
  \citenamefont {Xie}}]{zhang_bulk-boundary_2023}%
  \BibitemOpen
  \bibfield  {author} {\bibinfo {author} {\bibfnamefont {Z.-Q.}\ \bibnamefont
  {Zhang}}, \bibinfo {author} {\bibfnamefont {H.}~\bibnamefont {Liu}}, \bibinfo
  {author} {\bibfnamefont {H.}~\bibnamefont {Liu}}, \bibinfo {author}
  {\bibfnamefont {H.}~\bibnamefont {Jiang}},\ and\ \bibinfo {author}
  {\bibfnamefont {X.}~\bibnamefont {Xie}},\ }\bibfield  {title} {\bibinfo
  {title} {Bulk-boundary correspondence in disordered non-{{Hermitian}}
  systems},\ }\href {https://doi.org/10.1016/j.scib.2023.01.002} {\bibfield
  {journal} {\bibinfo  {journal} {Sci. Bull.}\ }\textbf {\bibinfo {volume}
  {68}},\ \bibinfo {pages} {157} (\bibinfo {year} {2023})}\BibitemShut
  {NoStop}%
\bibitem [{\citenamefont {Li}\ \emph {et~al.}(2023)\citenamefont {Li},
  \citenamefont {Wang}, \citenamefont {Song},\ and\ \citenamefont
  {Wang}}]{li_scalefree_2023}%
  \BibitemOpen
  \bibfield  {author} {\bibinfo {author} {\bibfnamefont {B.}~\bibnamefont
  {Li}}, \bibinfo {author} {\bibfnamefont {H.-R.}\ \bibnamefont {Wang}},
  \bibinfo {author} {\bibfnamefont {F.}~\bibnamefont {Song}},\ and\ \bibinfo
  {author} {\bibfnamefont {Z.}~\bibnamefont {Wang}},\ }\bibfield  {title}
  {\bibinfo {title} {Scale-free localization and {{PT}} symmetry breaking from
  local non-{{Hermiticity}}},\ }\href
  {https://doi.org/10.1103/PhysRevB.108.L161409} {\bibfield  {journal}
  {\bibinfo  {journal} {Phys. Rev. B}\ }\textbf {\bibinfo {volume} {108}},\
  \bibinfo {pages} {L161409} (\bibinfo {year} {2023})}\BibitemShut {NoStop}%
\bibitem [{\citenamefont {Okuma}\ \emph {et~al.}(2020)\citenamefont {Okuma},
  \citenamefont {Kawabata}, \citenamefont {Shiozaki},\ and\ \citenamefont
  {Sato}}]{okuma_topological_2020}%
  \BibitemOpen
  \bibfield  {author} {\bibinfo {author} {\bibfnamefont {N.}~\bibnamefont
  {Okuma}}, \bibinfo {author} {\bibfnamefont {K.}~\bibnamefont {Kawabata}},
  \bibinfo {author} {\bibfnamefont {K.}~\bibnamefont {Shiozaki}},\ and\
  \bibinfo {author} {\bibfnamefont {M.}~\bibnamefont {Sato}},\ }\bibfield
  {title} {\bibinfo {title} {Topological origin of non-{{Hermitian}} skin
  effects},\ }\href {https://doi.org/10.1103/PhysRevLett.124.086801} {\bibfield
   {journal} {\bibinfo  {journal} {Phys. Rev. Lett.}\ }\textbf {\bibinfo
  {volume} {124}},\ \bibinfo {pages} {086801} (\bibinfo {year}
  {2020})}\BibitemShut {NoStop}%
\bibitem [{\citenamefont {Zhang}\ \emph
  {et~al.}(2020{\natexlab{b}})\citenamefont {Zhang}, \citenamefont {Yang},\
  and\ \citenamefont {Fang}}]{zhang_correspondence_2020}%
  \BibitemOpen
  \bibfield  {author} {\bibinfo {author} {\bibfnamefont {K.}~\bibnamefont
  {Zhang}}, \bibinfo {author} {\bibfnamefont {Z.}~\bibnamefont {Yang}},\ and\
  \bibinfo {author} {\bibfnamefont {C.}~\bibnamefont {Fang}},\ }\bibfield
  {title} {\bibinfo {title} {Correspondence between winding numbers and skin
  modes in non-{Hermitian} systems},\ }\href
  {https://doi.org/10.1103/PhysRevLett.125.126402} {\bibfield  {journal}
  {\bibinfo  {journal} {Phys. Rev. Lett.}\ }\textbf {\bibinfo {volume} {125}},\
  \bibinfo {pages} {126402} (\bibinfo {year} {2020}{\natexlab{b}})}\BibitemShut
  {NoStop}%
\bibitem [{\citenamefont {Mandal}\ and\ \citenamefont
  {Banerjee}(2022)}]{mandal2022surface}%
  \BibitemOpen
  \bibfield  {author} {\bibinfo {author} {\bibfnamefont {D.}~\bibnamefont
  {Mandal}}\ and\ \bibinfo {author} {\bibfnamefont {S.}~\bibnamefont
  {Banerjee}},\ }\bibfield  {title} {\bibinfo {title} {Surface acoustic wave
  {(SAW)} sensors: Physics, materials, and applications},\ }\bibfield
  {journal} {\bibinfo  {journal} {Sensors}\ }\textbf {\bibinfo {volume} {22}},\
  \href {https://doi.org/https://doi.org/10.3390/s22030820}
  {https://doi.org/10.3390/s22030820} (\bibinfo {year} {2022})\BibitemShut
  {NoStop}%
\end{thebibliography}

%

\begin{acknowledgements}
B.-B.W., Y.G., H.-X.S., Q.-R.S. and S.-Q.Y. are supported by the National Natural Science Foundation of China (Grants No. 12274183, 12174159 and 51976079), the National Key R\&D Program of China (Grant No. 2020YFC1512403), the Research Project of State Key Laboratory of Mechanical System and Vibration (Grant No. MSV202201) and the Postgraduate Research and Practice Innovation Program of Jiangsu Province (Grant No. KYCX22\_3603). Z.C., and B.Z. are supported by the Singapore Ministry of Education Academic Research Fund Tier 2 (Grant No. MOE-T2EP50123-0007) and Singapore National Research Foundation Competitive Research Program (Grant No. NRF-CRP23-2019-0007). H.X. acknowledges support from the National Natural Science Foundation of China (Grant No. 62401491) and the Chinese University of Hong Kong (Grants No. 4937205 and 4937206).
\end{acknowledgements}

\bigskip
\noindent{\large{\bf{Author contributions}}}

\noindent H.-X.S., H.X. and B.Z. conceived the idea. Z.C. and B.-B.W. performed theoretical analysis and numerical calculations. B.-B.W., Y.G. and H.-X.S. designed the sample and experiment. B.-B.W., H.-Y.Z., K.-Q.Z., Q.-R.S. and H.-X.S. conducted the experiment. B.-B.W., Z.C., S.-Q.Y., H.-X.S., H.X. and B.Z. wrote the manuscript with input from all authors. H.-X.S., H.X. and B.Z. supervised the whole project.

\bigskip
\noindent{\large{\bf{Competing interests}}}

\noindent The authors declare no competing interests.

\end{document}